\documentclass[10pt, conference, compsocconf]{IEEEtran}
\ifCLASSINFOpdf
\else
\fi
\usepackage{mathptm}
\usepackage[utf8]{inputenc}
\usepackage[T1]{fontenc}
\usepackage{babel}
\usepackage{textcomp}
\usepackage{type1cm}
\usepackage{amsmath}
\usepackage{amsfonts}
\usepackage{subcaption}
\usepackage{booktabs}
\usepackage{listings}
\usepackage{graphicx}
\let\labelindent\relax
\usepackage{enumitem}
\usepackage{color}

\begin{document}
%
\title{Autotuning Benchmarking Techniques:\\A Roofline Model Case Study}


\author{
\IEEEauthorblockN{Jacob O. Tørring, Jan Christian Meyer, Anne C. Elster}
\IEEEauthorblockA{Department of Computer Science\\
Norwegian University of Science and Technology (NTNU)\\
Trondheim, Norway\\
Email: \{jacob.torring, jan.christian.meyer, elster\}@ntnu.no}
}

\maketitle

\begin{abstract}

Peak performance metrics published by vendors often do not  correspond to what can be achieved in practice. It is therefore of great interest to do extensive benchmarking on core applications and library routines. Since DGEMM is one of the most used in compute-intensive numerical codes, it is typically highly vendor optimized and of great interest for empirical benchmarks.

In this paper we show how to build a novel tool that autotunes the benchmarking process for the Roofline model. Our novel approach can efficiently and reliably find optimal configurations for any target hardware. Results of our tool on a range of hardware architectures and comparisons to theoretical peak performance are included.

Our tool autotunes the benchmarks for the target architecture by deciding the optimal parameters through state space reductions and exhaustive search. Our core idea includes calculating the confidence interval using the variance and mean and comparing it against the current optimum solution. We can then terminate the evaluation process early if the confidence interval’s  maximum  is  lower  than  the  current  optimum solution. This dynamic approach yields a search time improvement of up to 116.33x for the DGEMM benchmarking process compared to a traditional fixed sample-size methodology. Our tool produces the same benchmarking result with an error of less than 2\% for each of the optimization techniques we apply, while providing a great reduction in search time. We compare these results against hand-tuned benchmarking parameters. Results from the memory-intensive TRIAD benchmark, and some ideas for future directions are also included.
\end{abstract}


%
\IEEEpeerreviewmaketitle

\section{Introduction}\label{sec:introduction}

Knowing how much performance can be achieved on a given architecture is of great interests to both computing centers and their users.
Ideally, one should be able to model the performance of a computing system, and use it to estimate the performance of a given application. 
However, this is both challenging and likely requires a bottom-up approach to realistically model software and hardware interactions. Meyer and Elster~\cite{meyer_performance_2010} did this by composing system models from simpler, linear models, which allowed parts of the analysis to be automated. They associated empirically benchmarked platform performance metrics with the core elements in a variant of bulk-synchronous execution. However, good empirical benchmarking is still required to verify results.

For instance, the theoretical peak performance is often available in hardware specifications from vendors. However, these specifications have to be found manually and do not necessarily correspond to \emph{practical} peak performance. In this paper, we therefore present an automated procedure using benchmarks and autotuning to calculate the necessary parameters for the Roofline model.

The Roofline model was developed by Williams \emph{et al.} in 2009 as a visual performance model that would be easy to understand~\cite{williams_roofline_2009}. This model enables programmers to get necessary insights to improve floating-point performance in their parallel software, or selecting ideal hardware architectures for the software's characteristics. 

We characterize the software or hardware in the Roofline model using \emph{Operational Intensity}($I$), a measure of operations per byte of memory traffic.

High intensity benchmarks set an upper bound for floating point computation, giving us a practical peak compute performance. One of the most popular such benchmarks is the Double-precision General Matrix Multiplication(DGEMM), since it performs many operations per byte. Because of this, many numerical problems are cast in terms of DGEMM, and the vendors typically hand-tune this library routine to show off peak performance.

Low intensity benchmarks, on the other hand, are bounded by the performance of the memory subsystem that the benchmark targets. 
We thus also include the TRIAD benchmark from STREAM~\cite{stream_mccalpin_2007} to characterize low intensity workloads. This benchmark combines a vector product with a scaling of one vector, resulting in an operational intensity $I = \frac{1}{12} FLOP/byte$. By varying the size of these vectors we can fit the data into the DRAM or L3 cache and measure the respective performance. We can generate a Roofline graph automatically by using the results from these benchmarks, without needing specifications from vendors.

Both the DGEMM benchmark and the TRIAD benchmark are parallelized using OpenMP~\cite{dagum_openmp_1998}. OpenMP is an explicit programming model that supports shared-memory multiprocessing. DGEMM is implemented using OpenMP as part of the BLAS libraries used in this tool. For the TRIAD benchmark we use OpenMP explicitly to parallelize our core evaluation loop.

Benchmark parameters must be carefully tuned to extract the maximum performance from the target system. We therefore apply autotuning techniques to find the optimal parameters for each of these benchmarks.

Autotuning has been used to optimize programs for their target hardware, most notably the ATLAS~\cite{hall_automatic_1995} and FFTW~\cite{frigo_fftw_1998} projects.
The definition and reduction of the \emph{search space} is critical for autotuning. Constraint specification can have great impact on an auto-tuner's performance. We will therefore clearly construct the search space of our auto-tuner.

To reduce the search time further we implement a novel sample evaluation process that dynamically adapts itself to the variance of the sample and the current best known solution. We calculate the sample mean and variance using Welford's online variance algorithm~\cite{welford_note_1962}. Using this variance and mean we compute the respective confidence intervals. By comparing the confidence interval of our currently evaluating sample against the current optimum solution, we can terminate the evaluation process early if the confidence interval's maximum is lower than the current optimum solution. This approach gives a high confidence that the measurements have converged sufficiently to decide on the optimum solution. As we will show, this novel approach yields the same results as a traditional fixed sample-size approach, yet provides up to 116.33x performance increase in search time.

The rest of this paper is structured as follows: Section~\ref{sec:roofline} gives a detailed introduction to the Roofline model and discusses related work. Section~\ref{sec:benchmarking} describes how we constructed the benchmarks to allow for autotuning, while Section~\ref{sec:auto-tuning} introduces our autotuning approach. Section~\ref{sec:experimental-setup} details our experimental setup and Section~\ref{sec:results} presents our results and discusses their significance. Lastly, Section~\ref{sec:future-work-conclusion} concludes the article and suggests future work.

\section{Roofline Model and Related Work}\label{sec:roofline}
In this article we will be referring to Operational Intensity 
$I$ as the ratio between the work $W$ and the memory traffic $Q$. This expresses the number of bytes used per FLOP of work and is denoted FLOP/byte, as used by Ilic \emph{et al.}~\cite{ilic_cache-aware_2014}.  

\begin{equation}
    I = \frac{W}{Q}
\end{equation}

\begin{figure*}[t]
    \centering
    \includegraphics[width=\linewidth]{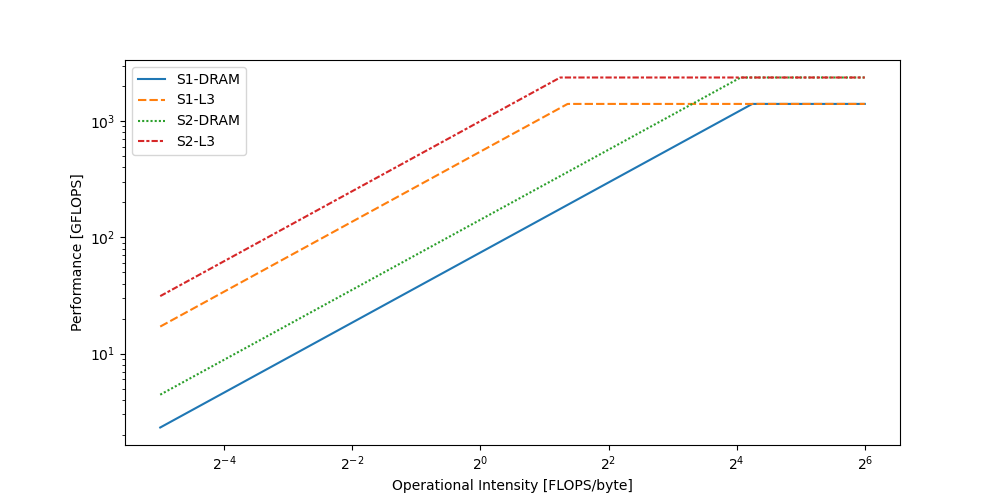}
    \caption{Roofline model example with four memory subsystems and two compute configurations. Single-socket DRAM, single-socket L3 Cache, dual-socket DRAM and dual-socket L3 cache.}
    \label{fig:example}
\end{figure*}

Given peak compute performance $F_p$ for a system and peak memory bandwidth $B_\alpha$ for a memory subsystem $\alpha$, we can calculate the function for the roofline graph using Eq.~\ref{eq:func}. 
\begin{equation}\label{eq:func}
    F_\alpha(I) = \min\{ B_\alpha \cdot I, F_p \}
\end{equation}

The axes of the roofline graph are logarithmic with the Operational Intensity $I$ on the X-axis and the corresponding performance of the system in GFLOP/s on the Y-axis. We present an example of such a graph in Fig~\ref{fig:example}.

The most relevant related work is the proprietary solution from Intel in their Intel Advisor tool~\cite{marques_performance_2017}. The Advisor provides both practical and theoretical upper bounds for the performance of an Intel chip. The main drawbacks of this tool are the lack of an open-source license, and the restriction of the tool to only support Intel products.

Ilic and Denoyelle have developed a toolset to create Cache-aware Roofline Models (CARMs)~\cite{ilic_cache-aware_2014, denoyelle_automatic_2016}, however, this toolset is based on microbenchmarks that use hardware specific details and are therefore not portable between different architectures. Recent works by Marques et al.~\cite{marques_application-driven_2020} still rely heavily on hand-coded microbenchmarks that target a specific architecture to extract the maximum performance from the machine. This has the significant limitation of restricting portability between systems.

\section{Benchmarking and Related Work}\label{sec:benchmarking}
In this section we will describe how we constructed the benchmarks necessary for the roofline model.
Both a benchmark with high Operational Intensity and a benchmark with low Intensity, are needed. To represent these respective categories, we chose the vendor's BLAS implementation of DGEMM, and an independently developed, portable TRIAD implementation.

There has been significant research into how to benchmark and compare results using statistically rigorous methods. Georges \emph{et al.}~\cite{georges_statistically_2007} separates the evaluation process into VM invocation-level and compilation-level repetitions. These layers provide different characteristics for how the performance varies. We therefore design our experiments to measure both levels of repetitions. 
The authors recommend assuming a Normal distribution when the number of samples $n$ satisfies $n\geq30$ and to check overlapping confidence intervals when comparing two samples.
They also discuss the idea of steady-state performance and how a program should ideally converge towards a steady-state performance over time. However, in some cases this never happens, and the performance continues to change indefinitely. 
The authors use the Coefficient of Variation(CoV) to determine when a program has reached steady-state. 
They also suggest terminating the evaluation process when the $\alpha$ confidence interval has converged to values that are $\pm1\%$ of the mean value. We use the same technique in our implementation.

Like Georges \emph{et al.}, Kalibera \emph{et al.}~\cite{kalibera_rigorous_2013} also has a concept of iteration-level repetition and execution-level repetitions. Iteration-level repetitions are repeated until an "independent state" is reached. If the benchmark does not reach an independent state in a reasonable amount of time, then they choose the same $i$th iteration of each execution repetition.

Kalibera \emph{et al.}~\cite{kalibera_quantifying_2020} builds upon their earlier work~\cite{kalibera_rigorous_2013} by providing a thorough investigation into handling uncertainty when benchmarking programs.
Hoefler \emph{et al.}~\cite{hoefler_scientific_2015} also give twelve recommendations for how to benchmark and report performance results, including the importance of confidence intervals for non-deterministic performance.
 
In this paper we will show how even a simple implementation of confidence intervals for benchmarking can be used to create optimizations that greatly increase autotuning performance.

\subsection{DGEMM}
The DGEMM benchmark consists of computing Eq.~\ref{eq:dgemm}, with the dimensions of matrix A being defined by $n\times k$, matrix B defined by $k\times m$ and matrix C defined by $n\times m$. For our benchmark we set $\alpha = 1.0$ and $\beta = 0.0$.
In addition to the matrix dimensions, the user can also provide a range of variables to determine \emph{the stop condition} which determines when the evaluation ends. This will be further described in Sec.~\ref{subsec:evaluation-budget}.

\begin{equation}\label{eq:dgemm}
    C \leftarrow \alpha AB + \beta C
\end{equation}
The DGEMM benchmark is thus composed of four main parts: input handling, test matrix initialization, inner iteration loop which repeatedly calculates Eq.~\ref{eq:dgemm}, and the stop condition evaluation. The program is also repeatedly invoked and evaluated in an outer invocation loop to stabilize the results.

We use the OpenBLAS and MKL BLAS libraries. Their parallelizations of the DGEMM include using OpenMP~\cite{dagum_openmp_1998}. The OpenMP core allocation policy determines how the problem is divided and scheduled to the available cores. For the DGEMM problem we want to keep the data as close to the processing cores as possible. In this case \lstinline{KMP_AFFINITY=close} will ensure that our cores are scheduled sequentially from ID 0 to ID N, so that the first N/2 processes are only allocated to the first CPU in a two socket system. 
After the initialization we call \lstinline{cblas_dgemm()} to pre-heat the hardware before starting the measurements.

Our measurement loop consists of a call to the DGEMM implementation of the target vendor's BLAS library. We record the time using {\tt{gettimeofday}} before and after the calls.
The number of FLOPS computed is calculated by taking the total number of floating points necessary to compute the DGEMM divided by the elapsed time. This value is then used to adjust the mean and evaluate the stop condition.

We also need to find the matrix dimensions $n, m, k$ that maximize the FLOPS in our \lstinline{cblas_dgemm} call. The FLOPS count vary greatly between architectures as they are heavily associated with available cache sizes, SIMD instructions, etc.

As we will see in Sec.~\ref{sec:results}, a poor choice of matrix dimensions for DGEMM, will lead to performance as low as 52.08\% of the peak theoretical performance. The relative performance increase is therefore great compared to a better selection of dimensions, that can reach as high as 98.06\% of theoretical maximum. We will therefore autotune the matrix dimensions to find the configuration that maximizes the DGEMM performance for the target system.

\subsection{TRIAD}
The TRIAD kernel is memory-bound, and thus puts an upper limit on the memory performance for each of the available memory subsystems. Eq~\ref{eq:triad} describes the triad kernel, where $A$, $B$ and $C$ are vectors and $\gamma$ is some scalar. 
\begin{equation}\label{eq:triad}
    C \leftarrow A + \gamma B
\end{equation}
This results in 2 floating point operations being computed for each iteration. In the case that the vectors represents an array of double precision floating points, these operations load and save a total of 3 doubles, or 24 bytes, per iteration. This results in a very low Operational Intensity of $\frac{2FLOP}{24byte} = \frac{1}{12} FLOP/byte$.

One can evaluate different memory subsystems by changing the size of the vector $n$ so that it fits into DRAM or L3 cache. Even with the high bandwidths of L3 cache, this kernel will still be memory-bound for most hardware systems.

We used OpenMP to parallelize the kernel, with a static schedule. The block size was left to the default value of $N/cores$, so as to evenly divide the vectors across all available cores. This leads to a benchmark where all available memory subsystems are stressed evenly, so that we can record the assumed practical maximum performance.

For our TRIAD benchmark we want to maximize bandwidth. By using \lstinline{KMP_AFFINITY=spread}, we spread out the threads across all of the available sockets, so that we can maximize the load on the DRAM memory subsystem. However, when the core count is equal to cores/sockets, the most realistic memory bandwidth is captured by running \lstinline{KMP_AFFINITY=close} as this will only use the memory channels of the currently evaluating sockets. If we had distributed the threads and data across multiple sockets, the cores would have access to an aggregate bandwidth of all the sockets.

\subsection{Evaluation Budget}\label{subsec:evaluation-budget}
\begin{figure}
    \centering
    \includegraphics[width=\linewidth]{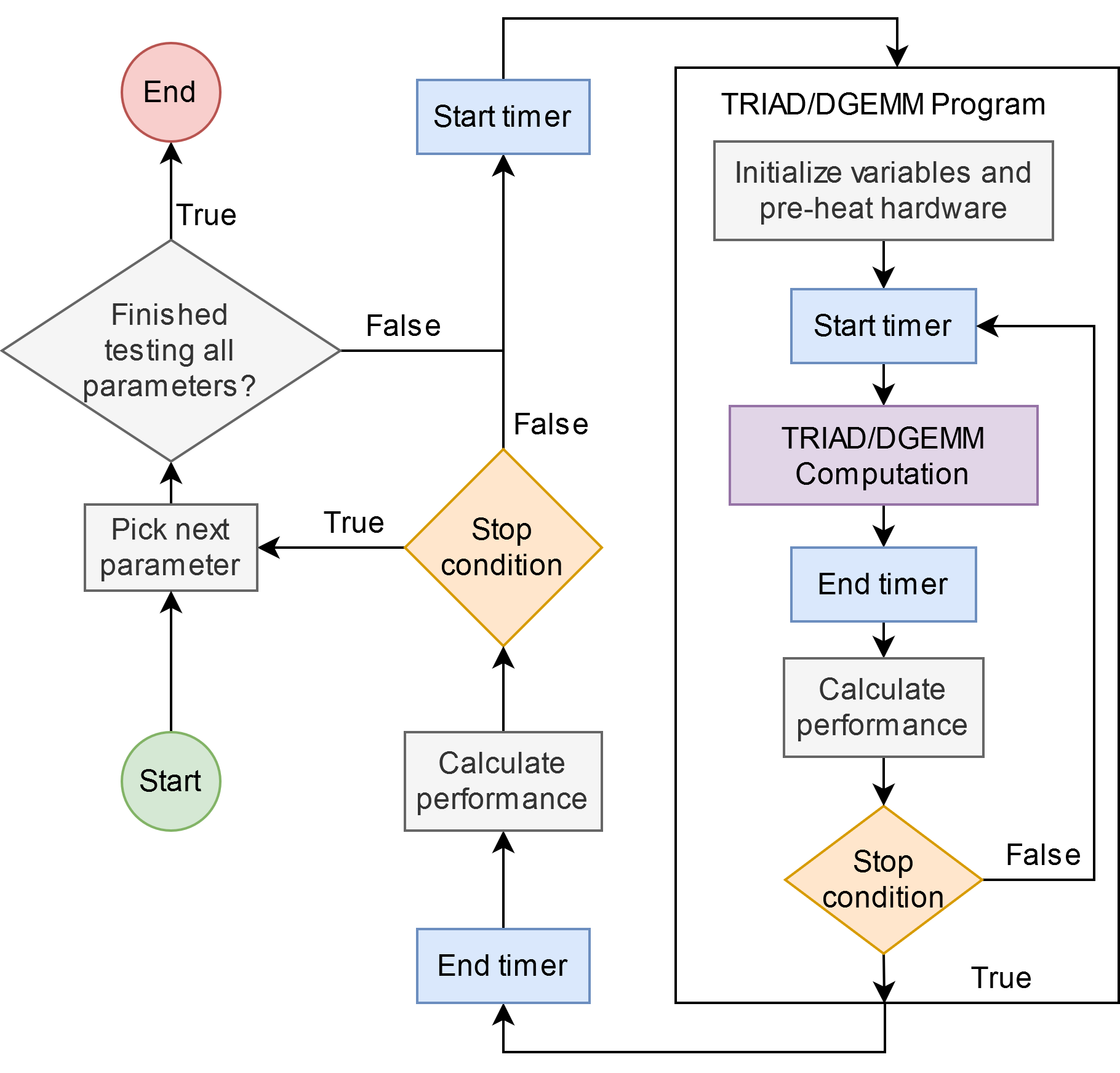}
    \caption{The autotuning benchmarking process, including the inner iteration loop and the outer invocation loop.}
    \label{fig:auto-tuning}
\end{figure}

The benchmarking process includes evaluation loops that repeatedly measure the DGEMM/TRIAD operations until one of four different stop conditions are met (Fig.~\ref{fig:auto-tuning}). We define the inner evaluation loop as the \emph{iteration loop} and the outer loop that executes the benchmark programs the \emph{invocation loop}. This section describes the four stop conditions that terminate this benchmarking process.

\subsubsection{Max time}
For each iteration of the innermost iteration loop, we record the time it took to perform the DGEMM or TRIAD operation and store the result. This elapsed time is then accumulated in a total time that we can use to ensure the benchmark only runs for a limited amount of time. The user can optionally set a maximum time threshold using the $-t$ flag or leave the benchmark at its default value.

\subsubsection{Max count}
The loop count is also accumulated. It thus provides statistics about how many iterations the benchmark required to reach a necessary confidence interval, or can be set as a maximum threshold. This count threshold is necessary as some configurations for the benchmark can get stuck in local optima, where the variance is high and the confidence interval converges slowly. According to our empirical results, these scenarios are rarely the top performing configurations. Therefore, setting an upper bound to the loop count can cut off the evaluation of such high variance configurations.

\subsubsection{Confidence interval of mean}
The iteration loop is evaluated a specified number of times $x$ in order to gain \emph{confidence} of how precise our result is. Instead of selecting an arbitrary loop count for all configurations, we have instead developed an approach where we automatically compute the \emph{confidence intervals} of our results. This enables our benchmark to run for only as long as necessary to achieve a certain precision. This avoids the pitfalls of running too few iterations for a high variance configuration, and it can terminate earlier than a fixed iteration if the variance is low. 
To compute the confidence interval we need to keep track of the mean and variance. To avoid storing each result explicitly to compute the sample variance, we use Welford's online variance algorithm~\cite{welford_note_1962}.

\begin{equation}\label{eq:sample-variance}
    S^2 = \frac{\sum_{i=1}^k (x_i-\bar{x})^2}{n-1} = \frac{C}{n-1},\quad \bar{x} = \sum_{i=1}^k x_i/k
\end{equation}

To calculate the sample variance in Eq.~\ref{eq:sample-variance} we need the corrected sum of squares $C$. We can calculate this iteratively using the following recursive algorithm. After $n$ steps the corrected sum of of squares $C_n$, Eq.~\ref{eq:corrected-sum-definition}, can be calculated using the sample mean $m_n$ Eq~\ref{eq:sample-mean-n}, and iterative corrections to the corrected sum. From these definitions we get our base cases. For the full proof and derivation of these results we refer the reader to the article by Welford~\cite{welford_note_1962}. 

\begin{equation}\label{eq:sample-mean-n}
    m_n = \sum_{i=1}^n x_i/n = \frac{n-1}{n}m_{(n-1)} + \frac{1}{n}x_n, \quad m_1 = x_1
\end{equation}

\begin{equation}\label{eq:corrected-sum-definition}
\begin{split}
    C_n = \sum_{i=1}^n(x_i - m_n)^2 = \quad &C_{(n-1)} + (\frac{n-1}{n})(x_n - m_{(n-1)})^2, \\
    \quad & C_1 = x_1-m_1 = 0 \\
\end{split}
\end{equation}

Assuming a normal distribution, we can use this sample variance to calculate the confidence interval. We then alter our program loop to terminate when the 99\% confidence interval reaches a boundary within 1\% of our mean value. 

When the distribution of runtimes of our benchmarks is graphed, we find that the distribution is usually non-normal. One would therefore ideally avoid the normality assumption in calculation of the confidence interval. However, to the authors' knowledge, there are no easily available alternatives for computing confidence intervals efficiently online. Bootstrapping~\cite{efron_introduction_1993} has been used as a technique for non-parametric distributions to produce confidence intervals. However, we have not been able to find any widely used algorithms for computing these intervals online efficiently. Bootstrapping will thus require reiterating and resampling all of the results for each iteration of the autotuning. It was therefore deemed too computationally expensive for this tool.

\subsubsection{Upper-bound of CI vs currently best solution}
Given a confidence interval(CI) of the currently evaluating configuration, one can compare the upper-bound of this CI with the performance of the currently best configuration. If past configurations have outperformed the upper-bound of the CI of the currently evaluating configuration, there is a very low likelihood of the currently evaluating configuration outperforming the previously best configuration. If this is the case, it is considered safe to terminate the evaluation of the current solution and stop the benchmarking process for this configuration. By defining \emph{marg} as the difference between the mean value and the upper bound of the confidence interval, the proposed stop condition can be implemented using Listing~\ref{lst:upper-bound}.

By default the lower-bound for the number of iterations necessary to trigger this stop condition is only two iterations. However, in certain cases the performance of the evaluating configuration can increase substantially during the evaluation process as more iterations are performed. In this case it can be useful to increase this minimum count.
\vspace{0.1cm}
\begin{lstlisting}[language=C, caption={Conditional for loop break},label={lst:upper-bound}]
    if(     mean + marg < best 
        &&  count >= min_count) 
        return 1;
\end{lstlisting}

\section{Autotuning}\label{sec:auto-tuning}
For an auto-tuner to be as efficient as possible, we need to clearly define and constrain our \emph{search space}. One can then select the optimal search technique for the target search space.
We first evaluate the search space of the DGEMM benchmark, before evaluating the TRIAD benchmark. The full autotuning process can be seen in Fig.~\ref{fig:auto-tuning}.

\subsection{Search Space for DGEMM}
The autotuning parameters for DGEMM are constrained to the three dimensions of the matrices that are computed. We specify these three dimensions as $n$, $m$ and $k$. Initially we propose a search range with steps of power of 2 from $64$ to $4096$ for $n$ and $m$ and $2$ to $2048$ for $k$. 
This initial search space evaluates to the cardinality in Eq.~\ref{eq:initial-search-space}.
\begin{equation}\label{eq:initial-search-space}
    S = n\times m\times k,\quad |S| = 7\cdot7\cdot11 = 539
\end{equation}

The cardinality of this search space is low enough that it could theoretically be searched exhaustively, given that the runtime of a single configuration is short.
Despite this low cardinality, we perform a study to reduce the search space further. A smaller search space allows more time to evaluate each configuration, to ensure that each result is accurate.

We perform experiments to assess a Constraint Specification of m=n, which would reduce the cardinality of the search space significantly. In Intel's benchmarking~\cite{story_tips_2017} Hu and Story only evaluate matrices where $m=n=k$, and find the optimal for $m=n=k=1000$. However, we find that in most cases non-square matrices yield significantly higher performance compared to square matrices.

Through experiments we noticed that low values for $n$, $m$ and $k$ performed poorly, and so we narrowed the search range to higher values, from $512$ to $4096$ for $n$ and $m$ and $64$ to $2048$ for $k$. This reduces the cardinality to $4\cdot4\cdot6=96$.

In accordance to Intel's guidelines for DGEMM performance with MKL~\cite{story_tips_2017} we also adjusted the leading dimensions to be a multiple of 2, instead of powers of 2, i.e. 500,1000,2000,4000. 

\subsection{Search Space for TRIAD}
When autotuning the TRIAD kernel, we can adjust the the size of the vector $N$.
The main objective of autotuning these parameters is to find the peak memory bandwidth, corresponding to the L3 Cache. The search range therefore starts at $3KiB$, and ends at $768MiB$. 
We are only able to measure the DRAM and L3 cache, as lower levels are outside of the scope of this technique.

\subsection{Autotuning techniques}
For autotuning problems with low cardinality and low sample cost such as these ones, simple search techniques like random search or exhaustive search are often ideal. 
This is due to the relatively higher overhead of advanced autotuning techniques, compared with the effectiveness of gathering more samples using simpler techniques. More advanced techniques based on metaheuristic optimization or machine-learning might have been applicable if the search space had been larger.
However for our specific autotuning search space it is sufficient to use exhaustive search to search through all available configurations.

\section{Experimental Setup}\label{sec:experimental-setup}
\begin{table}[]
    \centering
    \caption{Auto-tuner configuration for the experiments}
    \begin{tabular}{cccc}
        \toprule
         Invocations & Iterations & Timeout & Error \\
         \midrule
         10 & 200 & 10s & 100 \\
         \bottomrule
    \end{tabular}
    \label{tab:autotuning-parameters}
\end{table}

The systems used in our experiments are listed in Table~\ref{tab:hardware}. They are part of the Idun~\cite{sjalander_epic_2020} cluster at NTNU.
\begin{table}[]
    \centering
    \caption{Hardware specification for the benchmarked systems.}
    \begin{tabular}{cccccccc}
        \toprule
        System & $Freq_{CPU}$ & Cores & $AVX_{Type}$ & $AVX_{Units}$\\ & $Freq_D$ & $Channels_D$ & $L3_{Size}$ & Sockets \\
        \midrule
        2650 v4     & 2.2GHz & 12 & AVX2 & 1 \\ 
                    & 2400MHz & 4 & 30 MB & 2\\
        2695 v4     & 2.1GHz & 18  & AVX2 & 1 \\  
                    & 2400MHz & 4 & 45 MB & 2\\
        Gold 6132   & 2.6GHz & 14 & AVX512 & 2 \\
                    & 2666MHz & 6 & 19.25MB & 2\\
        Gold 6148   & 2.4GHz & 20 & AVX512 & 2 \\ 
                    & 2666MHz & 6 & 31.75MB & 2\\ 
        \bottomrule
    \end{tabular}
    \label{tab:hardware}
\end{table}
Using these specifications we can compute the theoretical peak double precision performance and peak memory bandwidth.
The theoretical maximum FLOPS, $F_t$, for Intel CPUs is achieved by utilizing the AVX512 vector instructions~\cite{reinders_high_2015}. $F_t$ can therefore be calculated using Eq~\ref{eq:theoretical-flop}, where the performance of an AVX512 unit is given by Eq.~\ref{eq:avx-compute}. Similarly we 
compute the theoretical maximum bandwidth, $B_t$, using Eq.~\ref{eq:theoretical_bandwidth}.
\begin{equation}\label{eq:theoretical-flop}
    F_t = freq \cdot cores \cdot AVX_{type} \cdot AVX_{units}  \cdot CPUs
\end{equation}
\begin{equation}\label{eq:avx-compute}
\begin{split}
    AVX512_{DP} &= \frac{|Vector|\cdot ops/cycle}{|DP|} = \frac{512 bits \cdot 2 ops/cycle}{8 byte} \\
    &= 16 ops/cycle
\end{split}
\end{equation}

\begin{equation}\label{eq:theoretical_bandwidth}
    B_t = freq \cdot channels \cdot bytes/cycle 
\end{equation}

\begin{table}[]
    \centering
    \caption{Theoretical maximum double precision performance and DRAM memory bandwidth for each hardware system.}
    \begin{tabular}{crr}
        \toprule
        System & $F_t$ & $B_t$ \\
        \midrule
        2650v4 & 422.4 GFLOP/s & 76.8 GB/s \\
        2695v4 & 604.8 GFLOP/s & 76.8 GB/s \\
        Gold 6132 & 1164.8 GFLOP/s & 127.968 GB/s \\
        Gold 6148 & 1536 GFLOP/s & 127.968 GB/s \\
        \bottomrule
    \end{tabular}
    \label{tab:theoretical-peak-performance}
\end{table}

We measure the performance of our tool by the ratio of the recorded peak performance over the theoretical peak performance: $\frac{F_p}{F_t}$ for compute and $\frac{B_D}{B_t}$ for bandwidth. The theoretical peak performance can be found in Table~\ref{tab:theoretical-peak-performance}.

For our experiments the autotuning tool was configured with the parameters in Table~\ref{tab:autotuning-parameters}. This specifies that the inner evaluation loop can maximum run for 200 iterations, while the outer loop invokes the program 10 times. The maximum time threshold for each invocation is set to 10s for each configuration, and the invocation's stop condition will terminate when the boundaries of the 99\% confidence interval reaches $\pm1\%$ of the mean.

When testing our optimizations we refer to stop condition 3. from Sec.~\ref{subsec:evaluation-budget} as "Confidence" or "C". Stop condition 4 applied to the inner iteration loop is abbreviated "Inner" or "I" and for the outer invocation loop "Outer" or "O". We also show how the results are affected by the ordering of the search space, by reversing the exhaustive search, referred to as "Reverse" or "R".

Intel's MKL BLAS implementation provided substantially higher performance than OpenBLAS, so we chose this BLAS implementation for our experiments.
The experiments are run using SLURM, with exclusive access to the test nodes and Hyperthreading disabled. We were not able to disable clock frequency scaling, which might have affected the AVX512 performance as well as providing unstable performance results that affect the variance of the benchmarking process.

\section{Results and Discussion}\label{sec:results}
In this section we will present and discuss the results from our DGEMM benchmark and TRIAD benchmark. We will then present how our autotuning optimizations affect the accuracy of our benchmarking results.
\subsection{DGEMM results}
We will first compare our autotuning results with other available sources. The closest related work is the work by Hu and Story from Intel~\cite{story_tips_2017}. They optimized the Intel MKL DGEMM calls for the Intel Xeon Silver 4110 and found the maximum performance from $n=m=k=1000$, providing a peak performance of $559.93GFLOP/s$. Using the hardware specifications for the Intel Xeon Silver 4110 we can calculate the theoretical maximum performance.
\begin{equation}
    F_t = 2.1\cdot8\cdot32\cdot1\cdot2 = 1075.2 GFLOP/s
\end{equation}
Where the 32 multiplier for the $AVX_{type}$ is due to the use of single-precision floating point operations. This results in a utilization of $559.93/1075.2 = 52.08\%$ of peak performance. We do not have access to the same processor that Intel used, however, we will compare these results to four systems of the same generation and older generation architectures.

When we run the same dimensions as Intel's best result, $n=m=k=1000$ for our Intel Xeon Gold 6132 system we achieve a peak performance of $1297.48 GFLOP/s$ against a theoretical maximum of $2329.6 GFLOP/s$, which results in 55.69\% of theoretical peak. If we compare this to our autotuned configuration for the same hardware with dimensions $n=4000, m=512, k=128$ we get $1750.24 GFLOP/s$ or $75.13\%$ of peak performance.

\begin{figure}
    \centering
    \includegraphics[width=\linewidth]{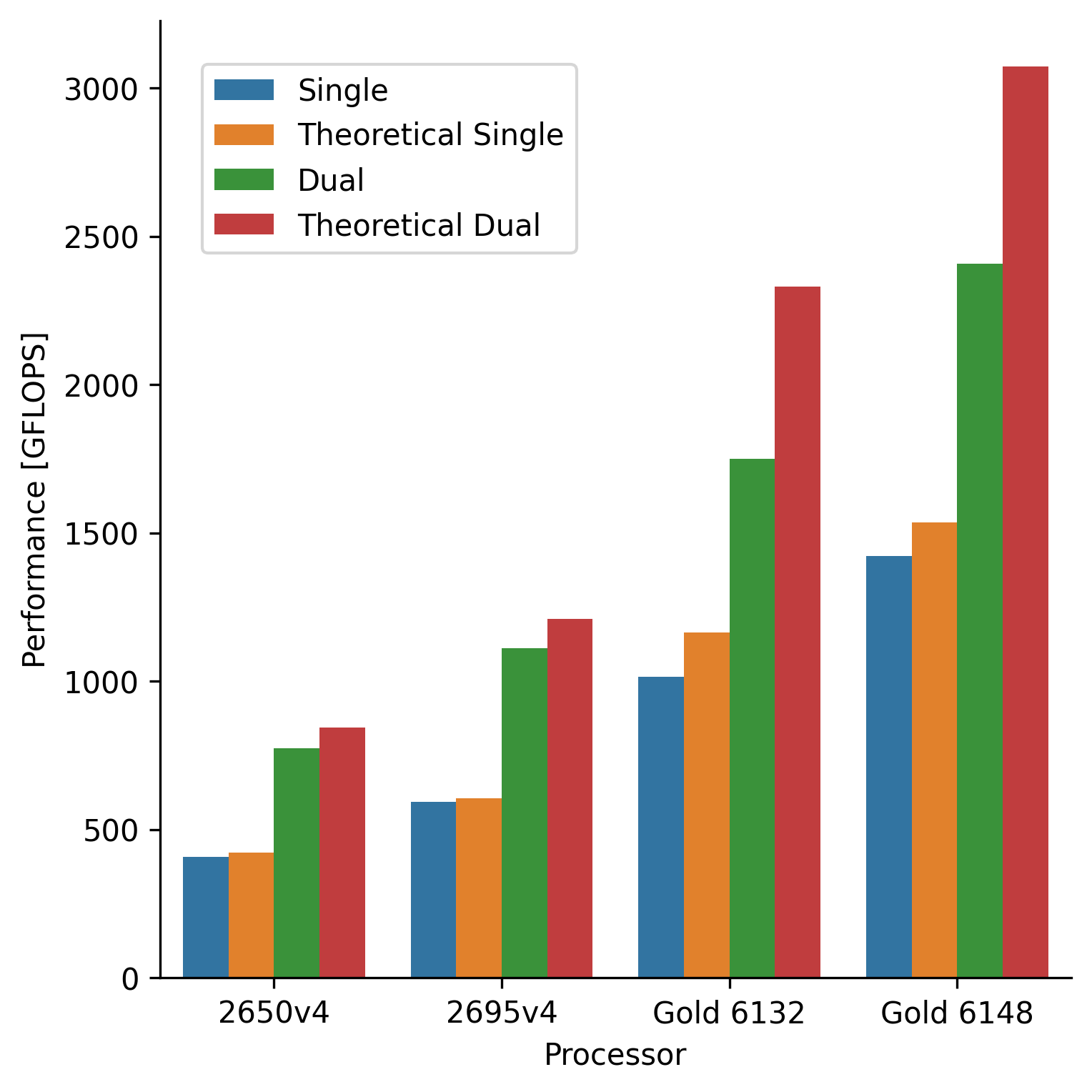}
    \caption{DGEMM compute performance vs. theoretical maximum performance for all systems and configurations.}
    \label{fig:dgemm-performance}
\end{figure}

The results from all of our DGEMM benchmarks are summarized in Table~\ref{tab:dgemm-results} and Fig.~\ref{fig:dgemm-performance}. We can observe that compute utilization is much higher for workloads that are restricted to a single socket, and that older generation hardware with only AVX2 units, had higher average utilization. We can also observe that all of our results greatly outperform the results from Intel's work~\cite{story_tips_2017}. The corresponding matrices for each of the optimal configurations is shown in Table~\ref{tab:dgemm-dims}. Here we can see that most hardware finds an optimal configuration with $k=128$ and that $n$ and $m$ varies depending on the hardware.

\begin{table}[]
    \centering
    \caption{Peak double-precision compute performance for each hardware system for single-socket and dual-socket configurations.}
    \begin{tabular}{crr}
        \toprule
        System      & $F_{S1}$     & $F_{S2}$ \\
        \midrule
        2650v4      & 408.71 (96.76\%)   & 773.51 (91.56\%) \\
        2695v4      & 593.06 (98.06\%)   & 1112.08 (91.93\%) \\
        Gold 6132   & 1015.68 (87.20\%)  & 1750.24 (75.13\%) \\
        Gold 6148   & 1422.24 (92.59\%)  & 2407.33 (78.36\%) \\
        \bottomrule
    \end{tabular}
    \label{tab:dgemm-results}
\end{table}

\begin{table}[]
    \centering
    \caption{Dimensions for the corresponding results from Table~\ref{tab:dgemm-results}.}
    \begin{tabular}{cll}
        \toprule
        System      & $F_{S1}$: n,m,k  & $F_{S2}$: n,m,k \\
        \midrule
        2650v4      & 1000,4096,128 & 2000,2048,64  \\
        2695v4      & 2000,4096,128 & 4000,2048,128 \\
        Gold 6132   & 1000,4096,128 & 4000,512,128  \\
        Gold 6148   & 4000,512,128  & 4000,1024,128 \\
        \bottomrule
    \end{tabular}
    \label{tab:dgemm-dims}
\end{table}

\begin{figure}
    \centering
    \includegraphics[width=\linewidth]{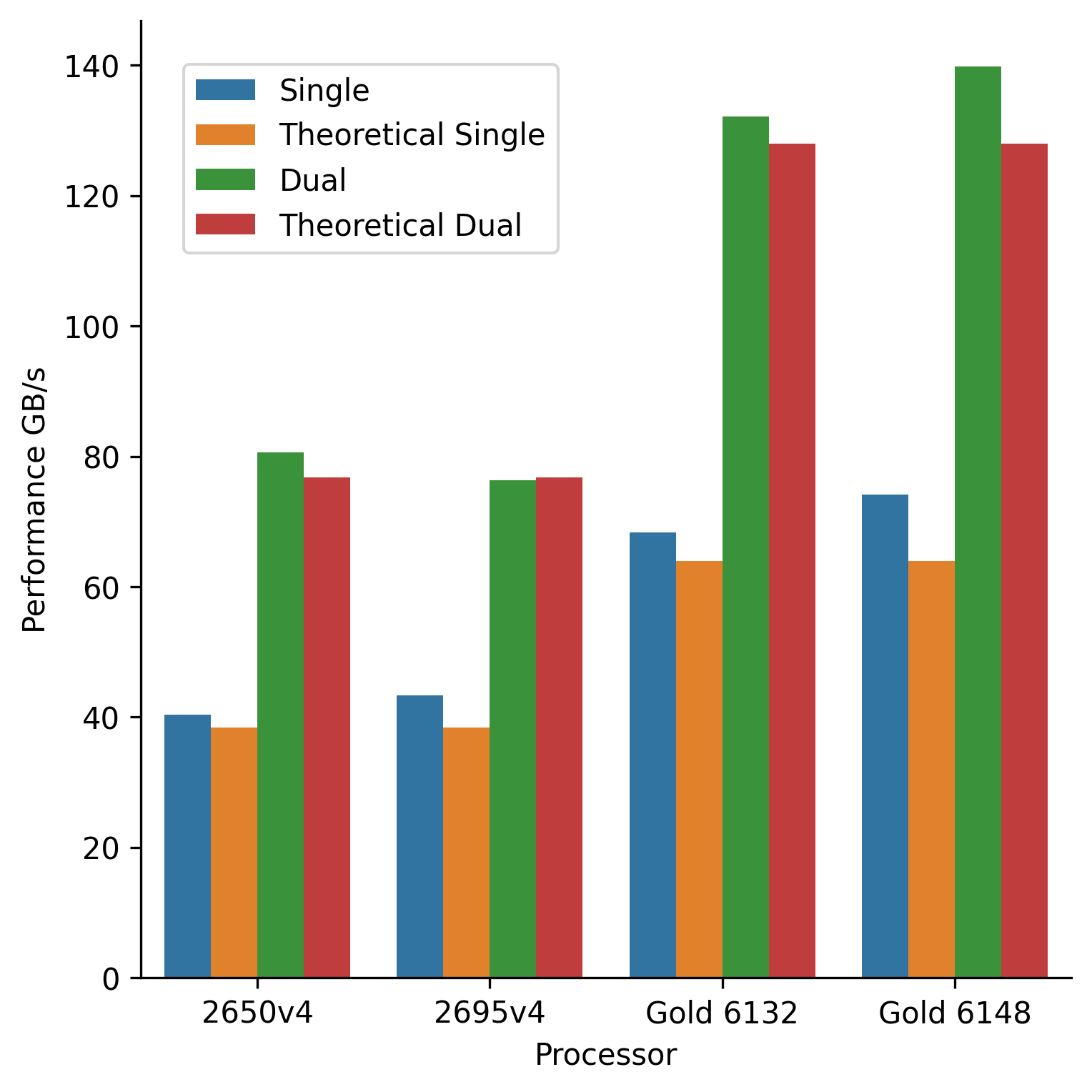}
    \caption{TRIAD memory performance vs. theoretical maximum performance for all systems and configurations.}
    \label{fig:TRIAD-performance}
\end{figure}

\subsection{TRIAD results}
For the memory results we find that the TRIAD kernel slightly overestimates the memory bandwidth of the system, we attribute this to noise from the L3 cache. 
We were unable to calculate the theoretical maximum bandwidth of the L3 cache and we therefore present the results as is. The results can be seen in Table~\ref{tab:memory-results} and Fig.~\ref{fig:TRIAD-performance}.

\begin{table}[]
    \centering
    \caption{Peak memory bandwidth for each hardware system and memory subsystem, for single-socket and dual-socket configurations.}
    \begin{tabular}{crrrr}
        \toprule
        System & $B_{DRAM,S1}$ & $B_{DRAM,S2}$ & $B_{L3,S1}$ & $B_{L3,S2}$ \\
        \midrule
        2650v4      &40.42(105.26\%) &80.65(105.01\%) & 256.07    & 452.05 \\
        2695v4      &43.29(112.73\%) &76.32(99.37\%)  & 371.41    & 661.68 \\
        Gold 6132   &68.32(106.78\%) &132.18(103.92\%) & 422.87    & 814.82 \\
        Gold 6148   &74.16(115.90\%) &139.80(109.25\%) & 547.11    & 1000.10 \\
        \bottomrule
    \end{tabular}
    \label{tab:memory-results}
\end{table}

\subsection{Optimizations}
\begin{figure*}
    \centering
    \includegraphics[width=\linewidth]{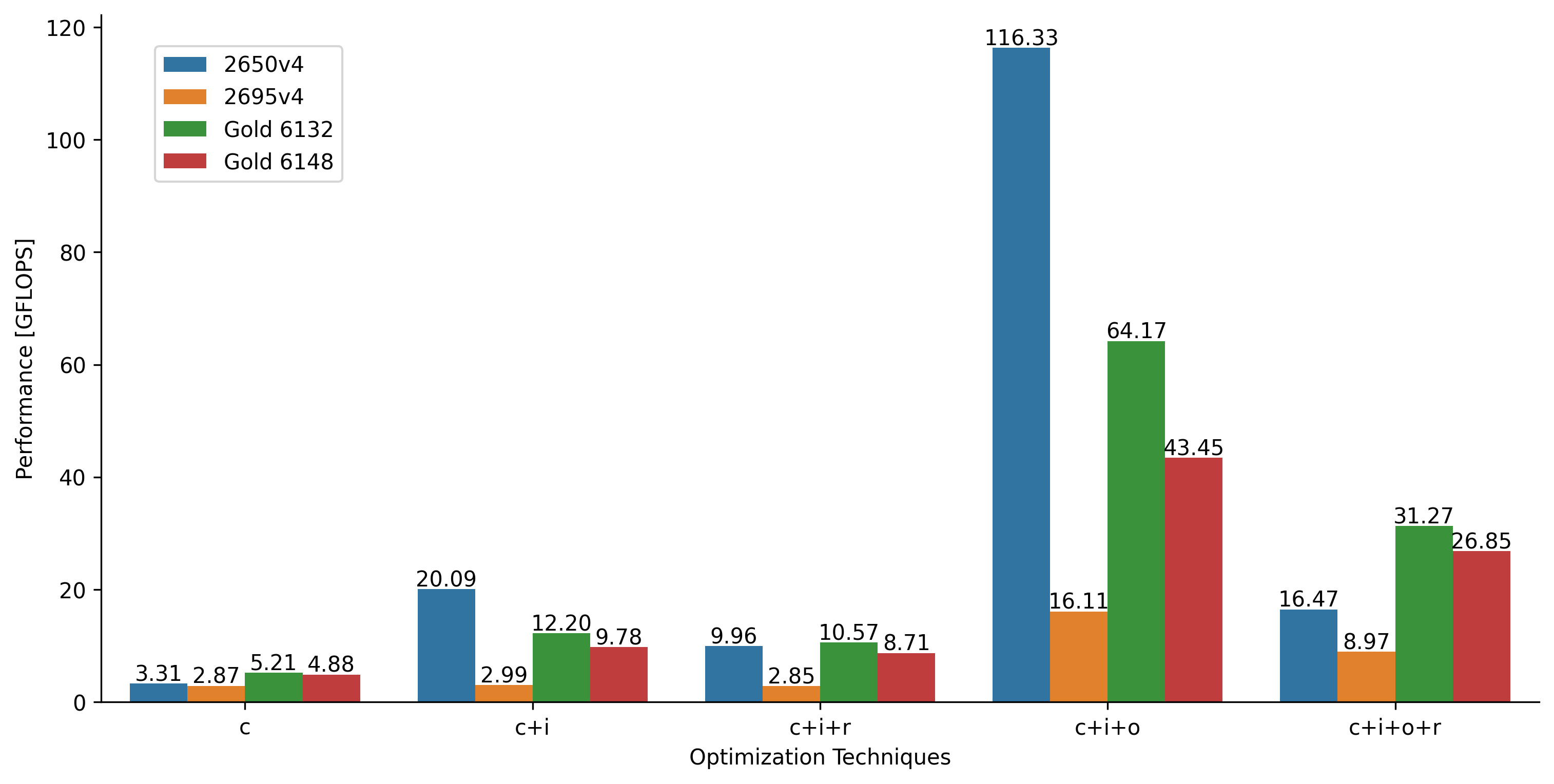}
    \caption{Performance increase over default solution for DGEMM benchmark on various hardware. Combinations of reversal of order, confidence interval, inner loop optimizations and outer loop optimizations.}
    \label{fig:dgemm-opts}
\end{figure*}

We experimented with how the optimizations affected benchmark results, and the total runtime of each benchmark under each optimization. The results are presented in Tables~\ref{tab:2650-opts}--\ref{tab:6148-opts} as well as Fig.~\ref{fig:dgemm-opts}.

All of the optimizations for the 2650v4, Gold 6132 and Gold 6148 found the same optimal matrix size as reported in Table~\ref{tab:dgemm-dims}. For 2695v4 the runs using the Upper-bound CI vs currently best solution optimization, with a lower count=100 also found the same results, except for the dual-socket benchmark with the c+i optimization. This specific optimization found a slightly worse-performing configuration: n=500, m=4096, k=1024.

\begin{table}[]
    \centering
    \caption{Iteration count for the time- and accuracy-tuned examples of each system}
    \begin{tabular}{ccc}
        \toprule
        System & $Iter_T$ & $Iter_A$ \\
        \midrule
        2650v4 & 7 & 20 \\
        2695v4 & 15 & 180 \\
        Gold 6132 & 18 & 180\\
        Gold 6148 & 30 & 150 \\
        \bottomrule
    \end{tabular}
    \label{tab:iterations}
\end{table}

For our hand-tuned experiments we set the outer invocation loop equal to one and tuned the inner iteration count to match the total runtime of our most optimized implementation. This provides the row "Hand-tuned Time". For "Hand-tuned Accuracy" we have tuned the inner iteration count upwards until the accuracy is comparable to our most optimized implementations. The number of iterations that we selected can be seen in Table~\ref{tab:iterations}.

From these results we see a significant improvement in search-time by applying the optimization techniques, and that the default settings preserve the benchmarking results for most systems. For the 2695v4 we add a minimum count=100 for the Upper-bound CI optimization, to ensure that the highest performing configurations were included. For all benchmarks the optimized autotuning process outperforms the traditional fixed sample size approach in both accuracy and performance.  

\begin{table}[]
    \centering
    \caption{Comparison of different evaluation optimizations for 2650v4}
    \begin{tabular}{lrrrrr}
        \toprule
        Technique & $F_{S1}$ Perf & $F_{S2}$ Perf & Time & Speedup  \\
        \midrule
        Default & 408.47 & 776.02 & 3435.73s & 1x \\
        Hand-tuned Time & 404.92 & 765.58 & 30.12s & 114.07x \\
        Hand-tuned Accuracy & 407.29 & 772.53 & 56.45s & 60.86x \\
        Single & 398.56 & 719.72 & 15.34s & 223.91x \\
        \midrule
        Confidence & 407.26 & 775.24 & 1039.03s & 3.31x \\ 
        C+Inner & 406.96 & 775.65 & 170.99s & 20.09x \\
        C+Inner+R & 406.99 & 774.92 & 344.92s & 9.96x \\
        C+I+Outer & 407.57 & 771.19 & 29.53s & 116.33x \\
        C+I+O+R & 406.84 & 775.08 & 208.61s & 16.47x \\
        \bottomrule
    \end{tabular}
    \label{tab:2650-opts}
\end{table}

\begin{table}[]
    \centering
    \caption{Comparison of evaluation optimizations for 2695v4.}
    \begin{tabular}{lrrrrr}
        \toprule
        Technique & $F_{S1}$ Perf & $F_{S2}$ Perf & Time & Speedup  \\
        \midrule
        Default & 590.47 & 1089.00 & 2531.58s & 1x \\
        Hand-tuned Time & 529.64 & 872.70 & 37.55s & 67.42x \\
        Hand-tuned Accuracy & 581.87 & 1064.24 & 237.84 & 10.64x \\
        Single & 436.35 & 634.16 & 19.24 & 131.58x \\
        \midrule
        Confidence & 587.26 & 1080.56 & 882.14s & 2.87x \\ 
        \midrule
        \multicolumn{5}{l}{Default optimizations} \\
        \midrule
        C+Inner & 467.48 & 931.81 & 201.34s & 12.57x \\
        C+Inner+R & 550.95 & 1018.42 & 338.02s & 7.49x \\
        C+I+Outer & 436.40 & 1011.02 & 35.94s & 70.44x \\
        C+I+O+R & 546.77 & 1013.77 & 174.81s & 14.48x \\
        \midrule
        \multicolumn{5}{l}{Minimum count=100 for stop condition 4 (See Sec.~\ref{subsec:evaluation-budget}.)} \\
        \midrule
        C+Inner & 587.10 & 1064.12 & 845.43s & 2.99x \\
        C+Inner+R & 587.05 & 1087.98 & 887.88s & 2.85x \\
        C+I+Outer & 587.11 & 1070.98 & 157.13s & 16.11x \\
        C+I+O+R & 586.77 & 1089.67 & 282.26s & 8.97x \\
        \bottomrule
    \end{tabular}
    \label{tab:2695-opts}
\end{table}

\begin{table}[]
    \centering
    \caption{Comparison of different evaluation optimizations for Gold 6132}
    \begin{tabular}{lrrrrr}
        \toprule
        Technique & $F_{S1}$ Perf & $F_{S2}$ Perf & Time & Speedup  \\
        \midrule
        Default & 1009.56 & 1756.06 & 1696.37s & 1x \\
        Hand-tuned Time & 992.36 & 1740.20 & 27.19s & 62.39x \\
        Hand-tuned Accuracy & 1005.34 & 1744.63 & 207.23s & 8.19x \\
        Single & 919.83 & 1401.98 & 12.78s & 132.74x \\
        \midrule
        Confidence & 1007.89 & 1748.46 & 325.34s & 5.21x \\ 
        C+Inner & 1007.27 & 1747.95 & 139.09s & 12.20x \\
        C+Inner+R & 1004.44 & 1745.84 & 160.50s & 10.57x \\
        C+I+Outer & 1006.51 & 1747.42 & 26.43s & 64.17x \\
        C+I+O+R & 1002.06 & 1745.60 & 54.26s & 31.27x \\
        \bottomrule
    \end{tabular}
    \label{tab:6132-opts}
\end{table}

\begin{table}[]
    \centering
    \caption{Comparison of different evaluation optimizations for Gold 6148}
    \begin{tabular}{lrrrrr}
        \toprule
        Technique & $F_{S1}$ Perf & $F_{S2}$ Perf & Time & Speedup  \\
        \midrule
        Default & 1408.14 & 2373.35 & 1409.28s & 1x \\
        Hand-tuned Time & 1342.37 & 2336.03 & 32.46s & 43.42x \\
        Hand-tuned Accuracy & 1405.02 & 2363.48 & 109.59s & 12.86x \\
        Single & 1221.08 & 1957.92 & 13.86s & 101.68x \\
        \midrule
        Confidence & 1403.46 & 2370.84 & 288.84s & 4.88x \\ 
        C+Inner & 1405.47 & 2368.21 & 144.08s & 9.78x \\
        C+Inner+R & 1402.60 & 2369.58 & 161.81s & 8.71x \\
        C+I+Outer & 1403.92 & 2373.57 & 32.43s & 43.45x \\
        C+I+O+R & 1403.13 & 2372.15 & 52.49s & 26.85x \\
        \bottomrule
    \end{tabular}
    \label{tab:6148-opts}
\end{table}

\begin{figure}
    \centering
    \includegraphics[width=\linewidth]{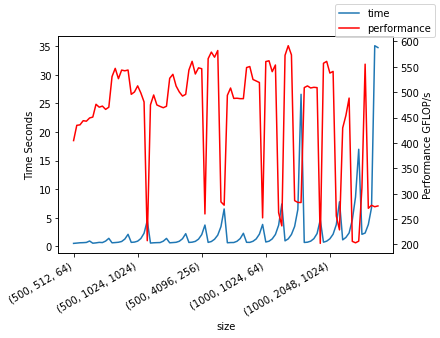}
    \caption{Time spent on each iteration and Performance as a function of matrix sizes.}
    \label{fig:time-vs-performance}
\end{figure}

We benchmarked the time and accuracy of running the benchmarks with a single invocation and a single iteration, where the results are presented as "Single" in the comparison tables.
We also present how the performance and time-consumption of each benchmark changes as the matrix sizes increase. While the performance peaks are spread out over the entire spectrum of matrix sizes, the time-consumption increases exponentially as the matrix sizes grows. A reversal of the parameter search therefore has significant impact upon the effectiveness of our benchmarking optimizations. Fig.~\ref{fig:time-vs-performance} shows this behavior.

\section{Conclusions and Future Work}\label{sec:future-work-conclusion}

Vendor-provided benchmarks often deviate from what can be achieved by empirical benchmarks.
However, picking and running benchmarks to properly characterize the system is time-intensive.
To address this, we presented a tool for automatically obtaining system Roofline models, using the DGEMM and TRIAD benchmarks to analyze hardware performance.

The results provided by our tool greatly outperformed the results provided by other available sources. For the 2695v4 single socket configuration the compute performance was as high as 98.06\% of theoretical maximum. The lowest performance was provided by the dual socket configuration of Gold 6132, providing 75.13\% of maximum performance. Older generation CPUs based on AVX2 had generally higher utilization of the available hardware, from 91.56\%-98.06\%, while newer generation systems with AVX512 provided 75.13\%-92.59\% of theoretical maximum. Single socket configurations had higher utilization of the compute hardware in general. These results indicated that for a high Operational Intensity problem such as DGEMM, the interconnect between CPUs bottleneck peak performance. The results also indicated that the AVX512 hardware in modern Intel CPUs is difficult to fully utilize.

By dynamically computing the confidence intervals of our evaluating configurations, we implemented autotuning optimizations that decreased search time by up to 116.33x, while providing the same DGEMM benchmarking results with an error of less than 2\%. We compared these results to hand-tuned autotuning parameters. 

The techniques presented in this paper are general autotuning benchmarking techniques that can be applied to any autotuning application.

For future work benchmarking more hardware such as L2 and L1 cache could be useful. Basing the stop conditions on other statistics, like the median, and basing the statistical tests on non-parametric statistics could also be useful. This should provide an even more robust solution. 
We could also change the data structures and how we compare the performance of difference configurations, to better handle configurations that achieve a high performance late into the iteration-count. This could improve the likelihood of not being skipped by the confidence interval. We can do this by having a time series of the performance of many configurations and make more advanced prediction towards when it is safe to terminate the evaluation process.  

\section*{Acknowledgements}
The authors would like to acknowledge the support of the High Performance Computing section at the NTNU IT Department, as well as the Department of Computer Science during the development of this project. The authors would also like to acknowledge Zawadi Berg Svela for their initial feedback on our paper.



%
\bibliographystyle{unsrt}
\bibliography{ms}
\end{document}